\documentclass{elsart}
\usepackage{xspace,amssymb,amsmath,helvet,graphics,setspace}
\begin{document}
\newcommand{\dee}{\,\mbox{d}}
\newcommand{\naive}{na\"{\i}ve }
\newcommand{\eg}{e.g.\xspace}
\newcommand{\ie}{i.e.\xspace}
\newcommand{\pdf}{pdf.\xspace}
\newcommand{\etc}{etc.\@\xspace}
\newcommand{\PhD}{Ph.D.\xspace}
\newcommand{\MSc}{M.Sc.\xspace}
\newcommand{\BA}{B.A.\xspace}
\newcommand{\MA}{M.A.\xspace}
\newcommand{\role}{r\^{o}le}
\def\bSig\mathbf{\Sigma}
\newcommand{\VS}{V\&S}
\newcommand{\tr}{\mbox{tr}}

\begin{frontmatter}
\title{Copulas from Order Statistics}
\address{Centre for Operational Research and Applied
Statistics, School of Business, University of Salford, UK}
\author{Rose Baker}
\ead{r.d.baker@salford.ac.uk}
\begin{abstract}
A new class of copulas based on order statistics was introduced
by Baker (2008). Here, further properties of the bivariate and
multivariate copulas are described, such as that of likelihood
ratio dominance (LRD), and further bivariate copulas are
introduced that generalize the earlier work. One of the new
copulas is an integral of a product of Bessel functions of
imaginary argument, and can attain the Fr\'{e}chet bound. The
use of these copulas for fitting data is described, and
illustrated with examples. It was found empirically that the
multivariate copulas previously proposed are not flexible enough to be
generally useful in data fitting, and further development is needed in
this area.
\end{abstract}
\end{frontmatter}
\section{Introduction}
The use of copulas has become  popular for modeling
multivariate data. Initially, the marginal distributions are fitted, using
the vast range of univariate models available, and the
dependence between variables is then modeled using a copula.
This approach is sometimes easier than seeking a `natural'
multivariate distribution derived from a probabilistic model,
because there may be no suitable multivariate distribution with
the required marginals.

Baker \cite{me} gave a class of bivariate and multivariate
copulas based on order-statistics, and this work seeks to `dig
deeper'. The ordering and some other properties of
these copulas are derived here, and the copulas are generalized into
further copulas in the bivariate case. Further experience is
also  gained in fitting distributions derived from the
bivariate and multivariate copulas to data.

First, we briefly recapitulate the essential concept of the
earlier work. Many topics covered there, such as the connection with
Bernstein polynomials \cite{lorenz} and the Farlie-Gumbel-Morgenstern (FGM)
distribution \cite{drouet}, are not repeated here.

The derivation of the class of distributions of interest is most
easily done by considering the generation of correlated random
variables from independent random variables $X$ and $Y$ with
distribution functions $F(x), G(y)$ and pdfs (where defined)
$f(x), g(y)$.  If $n$ sets of random variables $X_1 \dots X_n$
and $Y_1 \dots Y_n$ are sorted into order statistics
$X_{(1)}\dots X_{(n)}$ and $Y_{(1)} \dots Y_{(n)}$, they can be
paired off as $(X_{(1)}, Y_{(1)}), \dots (X_{(n)}, Y_{(n)})$,
and one such pair randomly selected. This scheme yields a pair
of dependent (positively correlated) random variables, the
Spearman (grade) correlation between them being $(n-1)/(n+1)$.
The marginal distributions of $X$ and $Y$ are still $F(x), G(y)$
respectively, because a randomly chosen order-statistic from a
distribution is just a random variable from that distribution.
We term the resulting bivariate distribution the `bivariate distribution of order
$n$'. The procedure also works in the general multivariate
case, when $n$ sets of $p$ random variables can be similarly
grouped.

In addition to random variable pairs being selected as
described, to be `in phase', they can be chosen to be in
antiphase, by pairing $X_{(k)}$ with $Y_{(n+1-k)}$, so giving a
negative grade correlation of $-(n-1)/(n+1)$. This is not
discussed further; to model negative correlations one simply replaces
$G(y)$ by $1-G(y)$ in the formulas. 

Baker \cite{me} obtained a one-parameter family of bivariate
copulas for a given $n$ by randomly choosing a pair from $(X_{(1)}, Y_{(1)}),
\dots (X_{(n)}, Y_{(n)})$  with probability $q$, and a pair from
$(X_1, Y_1) \ldots (X_n, Y_n)$ with probability $1-q$; these
random variables have grade correlation $q(n-1)/(n+1)$.
Equivalently, with probability $1-q$ we can choose $X$ and $Y$
randomly and independently from their $n$ order statistics
$X_{(1)} \ldots X_{(n)}$ etc; we could then say that $X$ and $Y$ are chosen from independent
cycles.  The resulting distributions are a mixture of the
distribution of order $n$ and the independent distribution. We
term them `mixture distributions of order $n$'. The device
of taking mixtures of copulas will be used later to derive new
copulas.

Some mathematical preliminaries are necessary: the distribution
function $F_{k,n}(x)$ of the $k$th of $n$ order statistics is
given by
\begin{equation}F_{k,n}(x)=\sum_{j=k}^n {n \choose j} F(x)^j(1-F(x))^{n-j}\label{eq:fkn}\end{equation}
(Stuart and Ord \cite{kands}).
The corresponding pdf if it exists is
\begin{equation}f_{k,n}(x)=n{n-1 \choose k-1}F(x)^{k-1}(1-F(x))^{n-k}f(x),\label{eq:fpdf}\end{equation}
and the bivariate distribution function of a random
order-statistic pair is
\begin{equation}H^{(n)}(x,y)=n^{-1}\sum_{k=1}^n F_{k,n}(x)G_{k,n}(y).\label{eq:hn}\end{equation}

The mixture distribution of order $n$ then has distribution
function 
\begin{eqnarray}H(x,y)&=&(1-q)F(x)G(y)+qn^{-1}\sum_{k=1}^n
F_{k,n}(x)G_{k,n}(y) \label{eq:plus}\\ 
&=&(1-q)H^{(1)}(x,y)+qH^{(n)}(x,y),\nonumber\end{eqnarray}
and where applicable, pdf
\begin{equation}h(x,y)=(1-q)f(x)g(y)+qn^{-1}\sum_{k=1}^n f_{k,n}(x)g_{k,n}(y).\label{eq:pdfp}\end{equation}
Note that the copula in fact has one continuous and one discrete
parameter.

There is no need to pair corresponding order statistics: in the most general
case the pair $(X_{(i)}, Y_{(j)})$ can be chosen with
probability $r_{ij}$, so that
\begin{equation}H(x,y)=\sum_{i=1}^n\sum_{j=1}^n r_{ij}F_{i,n}(x)G_{j,n}(y),\label{eq:gen}\end{equation}
where for the correct marginal distributions, we must have
\begin{equation}\sum_{i=1}^n r_{ij}=\sum_{j=1}^n r_{ij}=1/n.\label{eq:q}\end{equation}
The matrix $n{\mathbf r}$ is doubly stochastic, and ${\mathbf r}$
has $(n-1)^2$ independent elements.
Equation (\ref{eq:plus}) corresponds to the choice $r_{ij}=(1-q)/n^2+(q/n)\delta_{ij}$.

The Birkhoff-von Neumann theorem \cite{birk} states that the set of doubly
stochastic matrices of order $n$ is the convex hull of the
set of permutation matrices of order $n$, and  that the extreme
points of the set are the permutation matrices. Here, we can
view (\ref{eq:gen}) as a mixture of the $n!$ possible pairings
of the $X$ and $Y$ order statistics. The pairing chosen in
(\ref{eq:plus}) can achieve the largest grade
correlation for a given $n$, so the corresponding copula is of
special interest.
When considering copulas rather than bivariate distribution functions, it is convenient to define analogously to (\ref{eq:fkn})
the distribution functions of order statistics of $u=F(x), v=G(y)$ as
\[Q_{i,n}(u)=\sum_{l=i}^n B_{l,n}(u),\]
where $B_{i,n}(u)={n \choose i}u^i(1-u)^{n-i}$ is a Bernstein polynomial;
see Lorenz \cite{lorenz} for their mathematical description.
Then the copula corresponding to (\ref{eq:hn}) (the copula of
order $n$) is
\begin{equation}C(u,v)=n^{-1}\sum_{k=1}^n Q_{k,n}(u)Q_{k,n}(v) \label{eq:oldcop}\end{equation}
and for completeness, the mixture copula is
\begin{equation}C(u,v)=(1-q)uv+qn^{-1}\sum_{k=1}^n Q_{k,n}(u)Q_{k,n}(v). \label{eq:mixcop}\end{equation}

The material presented so far, except for the nomenclature and
the remarks on the Birkhoff-von Neumann theorem, was given in
\cite{me}. The essence of the earlier paper was the derivation
of the bivariate distribution of order $n$ (\ref{eq:hn}) and its
mixture with a distribution of order 1 to obtain the mixture
distribution of order $n$ (\ref{eq:plus}). This distribution
allows arbitrary correlations, and the corresponding copula
(\ref{eq:mixcop}) has the unusual feature of possessing one
continuous and one discrete parameter.

From this point on, new results are
presented. The mixture copula (\ref{eq:mixcop}) is of interest
in itself, and some further properties of it are derived, such
as its ordering properties. It is however convenient to start
from the more basic copula of order $n$ (\ref{eq:oldcop})
both when deriving properties of (\ref{eq:mixcop}) and when
deriving further copulas. In the next section, some further
properties of the bivariate copulas (\ref{eq:oldcop}) and
(\ref{eq:mixcop}) are given.
\section{Properties of the bivariate copula}
\subsection{Dependence Orderings}
The strongest ordering property is positive likelihood ratio
dependence (LRD), where $h(x_1,y_1)h(x_2,y_2) >
h(x_1,y_2)h(x_2,y_1)$ when $x_2 > x_1, y_2 > y_1$. The LRD
property implies all other quadrant dependence properties
(Nelsen, 2006 \cite{nelsen}). The FGM distribution is known to be LRD (eg
Drouet-Mari and Kotz, 2001 \cite{drouet}) and it is proved here that
(\ref{eq:hn}) is LRD. The general distribution (\ref{eq:gen}) 
is not.

Writing for brevity $F(x_1)=F_1, 1-F(x_1)=S_1, 1-G(y_1)=T_1$
etc, from (\ref{eq:hn}) we have that
\[\frac{h(x_1,y_1)h(x_2,y_2)-h(x_1,y_2)h(x_2,y_1)}{f(x_1)f(x_2)g(y_1)g(y_2)}=
n^2\sum_{i=1}^n\sum_{j=1}^n {n-1 \choose i-1}^2{n-1 \choose
j-1}^2 A_{ij}.\]
where
\[A_{ij}=\{(F_1G_1)^{i-1}(S_1T_1)^{n-i}(F_2G_2)^{j-1}(S_2T_2)^{n-j}-(F_1G_2)^{i-1}(S_1T_2)^{n-i}(F_2G_1)^{j-1}(S_2T_1)^{n-j}\}.\]
Since $A_{ii}=0$, the right-hand side can be rewritten as $n^2\sum_{i=1}^n\sum_{j=i+1}^n {n-1 \choose i-1}^2{n-1 \choose
j-1}^2 (A_{ij}+A_{ji})$. This can be factored into
\begin{equation}A_{ij}+A_{ji}=(F_1F_2G_1G_2)^{i-1}(S_1S_2T_1T_2)^{n-i}\{(F_2/S_2)^{j-i}-(F_1/S_1)^{j-i}\}\{(G_2/T_2)^{j-i}-(G_1/T_1)^{j-i}\}.\label{eq:brack}\end{equation}
Since $F_2 > F_1, G_2 > G_1$, then $F_2/S_2 > F_1/S_1, G_2/T_2 >
G_1/T_1$. As $j-i \ge 1$, each bracket of (\ref{eq:brack}) is
positive, and the LRD property follows. It follows
straightforwardly that mixture distributions derived from
(\ref{eq:hn}) such as (\ref{eq:plus}) are also LRD.
\subsection{Measures of Association}
The calculation of Kendall's tau for (\ref{eq:oldcop}) and
(\ref{eq:mixcop}) is given in
Baker \cite{me}. Blomqvist's medial coefficient or beta is another
widely used measure of association, given by $\beta=4C(1/2,1/2)-1$.
From (\ref{eq:hn}),
\[\beta=(4/n)(1/2)^{2n}\sum_{k=1}^n(\sum_{i=k}^n {n \choose i})^2-1.\]
This does not simplify much; it can also be written
\[\beta=(1/2)^{2n-2}\{{2n-1 \choose n-1}+2\sum_{i=1}^n {n-1
\choose i-1}\sum_{j > i}^n {n \choose j}\}-1.\]
There is an additional factor of $q$ for the copula (\ref{eq:mixcop}).

Note that at the median $(\tilde{x}, \tilde{y})$, $F(\tilde{x})=G(\tilde{y})=1/2$, and the pdf from (\ref{eq:hn}) takes a
simple form because the series can then be summed, to give
\[h(\tilde{x},\tilde{y})=f(\tilde{x})g(\tilde{y})(1/2)^{2(n-1)}n
{2n-2 \choose n-1}.\]
Since ${2m \choose m} < 2^{2m}/\sqrt{\pi m}$, we have that
$h(\tilde{x},\tilde{y}) < f(\tilde{x})g(\tilde{y})n/\sqrt{\pi(n-1)}$.

Gini's gamma is a coefficient of association that can be
expressed as $\gamma=4\int_0^1\{ C(u,u)+C(u,1-u)\} \dee u-2$
(Nelsen, \cite{nelsen}). For (\ref{eq:oldcop}) this gives after some algebra
\[\gamma=\frac{4}{n(2n+1)}\{\sum_{i=1}^n\sum_{j=1}^n (\min(i,j)+\min(i,n-j)){n
\choose i}{n \choose j}{2n \choose i+j}^{-1}\}-2.\]
There is again an additional factor of $q$ for the copula (\ref{eq:mixcop}).

Note that the Schweizer-Wolff sigma defined as
$\sigma=12\int\int|C(u,v)-uv|\dee u\dee v$ is numerically
identical to Spearman's rho for (\ref{eq:plus}), because it
possesses the PQD (positive quadrant dependence) property $C(u,v)
> uv$ as a consequence of the LRD property.

These dependence measures are shown in figure 1 plotted against
$n$. Dependence increases with $n$ and the Fr\'{e}chet bound is
attained as $n \rightarrow \infty$.

The coefficient of tail dependence (e.g. Joe \cite{joe}) is defined
in general as $\lambda=\lim_{p\rightarrow 0}\{Pr(y>y^*|x > x^*)$, where
$F(x^*)=1-p$, $F(y^*)=1-p$. From its definition, $0 \le \lambda
\le 1$, and $\lim_{p\rightarrow 0}\{Pr(y>y^*|x > x^*)=\lim_{p\rightarrow 0}\{Pr(x>x^*|y > y^*)$.
The distribution (\ref{eq:hn}) can be shown after some algebra
to yield $\lambda=0$, so that the random variables are asymptotically
independent. This property also holds for all finite mixture distributions.
\subsection{Symmetry}
The copula (\ref{eq:oldcop}) and its mixtures possess the radial
or reflective symmetry $C(1-u,1-v)=1-u-v+C(u,v)$, also seen for
example in the Frank and Plackett copulas.
All existing copulas seem to have the simpler symmetry property
$C(v,u)=C(u,v)$. Asymmetry between $X$ and $Y$
is usually handled by using different marginal distributions
$F(x)$ and $G(y)$, but there is no reason why the copula itself
should  be symmetric. Asymmetry can not occur for Archimedean
copulas, for which $C(v,u)=C(u,v)$ since
$\varphi(C(u,v))=\varphi(u)+\varphi(v)$, where $\varphi$ is a
(decreasing) function. Yet, as order statistics of $Y$ can be
paired with any permutation of order statistics of $X$, and
still give marginal distributions $F(x), G(y)$, it is
easy to construct copulas which do not have this symmetry.

For example, when $n=3$, the copula
\[C(u,v)=(1/3)\{Q_{1,3}(u)Q_{2,3}(v)+Q_{2,3}(u)Q_{3,3}(v)+Q_{3,3}(u)Q_{1,3}(v)\}\]
which can be written
\begin{eqnarray*}C(u,v)&=&(1/3)\{B_{1,3}(u)B_{2,3}(v)+B_{1,3}(u)B_{3,3}(v)+B_{2,3}(u)B_{2,3}(v)+2B_{2,3}(u)B_{3,3}(v)\\
\nonumber
&&+B_{3,3}(u)B_{1,3}(v)+2B_{3,3}(u)B_{2,3}(v)+3B_{3,3}(u)B_{3,3}(v)\}\end{eqnarray*}
is asymmetric.
It would
be interesting to devise tests of this symmetry and to see
whether such asymmetric copulas are ever needed in practice.
\subsection{Miscellaneous properties}
This short section covers four small points for completeness.
The hazard function $z(x,y)$ takes simple forms in the tails. When $F(x) \ll
1, G(y) \ll 1$, since the survival function
$S(x,y)=1-F(x)-G(y)+H(x,y)$, we have that $S(x,y) \sim 1$. Only
the $k=n$ term of the pdf from (\ref{eq:plus}) survives, and
this gives $z(x,y) \sim f(x)g(y)\{(1-q)+nq\}$. The
correlation between the random variables inflates the hazard. In the
right-hand tail, where $A(x)=1-F(x) \ll 1$, $B(x)=1-G(x) \ll 1$,
only the $j=n$ term survives from (\ref{eq:fkn}), for all $k$.
It follows that for $q > 0$, we have 
\[z(x,y) \sim \frac{f(x)g(y)\{nq+(1-q)\}}{q\{(1-F(x))+(1-G(y))\}}.\]
As the denominator can be much larger than $(1-F(x))(1-G(y))$,
the correlation between the random variables can decrease the hazard in the tail.

The pdf from (\ref{eq:hn}) can be written as a hypergeometric function:
\[\frac{h(x,y)}{f(x)g(y)}=n(F(x)G(y))^{n-1}~ _2F_1(1-n,1-n;1;\frac{(1-F(x))(1-G(y))}{F(x)G(y)}).\]

Median regression is the curve $y=\tilde{y}(x)$, where $\Pr(Y \le
\tilde{y}|X=x)=1/2$. This does not take any simple form for
these distributions.

Finally, the pdf from (\ref{eq:hn}) is proportional to the probability $p_{00}(2n)$
that a random walk in the plane returns to its start point after
$2n$ steps.  Given probabilities $p_1, p_2$ of moving left or
right, and probabilities $q_1, q_2$ of moving up or down, so
that $p_1+p_2+q_1+q_2=1$, we have that
\[p_{00}(2n)=\sum_{k=0}^n \frac{(2n)!}{k!^2(n-k)!^2}(p_1p_2)^k (q_1q_2)^{n-k},\]
so that 
\[\frac{h(x,y)}{f(x)g(y)}=\frac{2^{2n}n!(n-1)!}{(2n)!}p_{00}(2n-2),\]
where $FG=4p_1p_2$, $(1-F)(1-G)=4q_1q_2$, e.g. $p_1=F/2,
p_2=G/2, q_1=(1-F)/2, q_2=(1-G)/2$. At the median where
$F(x)=G(y)=1/2$, the random walk is symmetric, with
probability $1/4$ of moving in any direction.
\section{Further Bivariate models}
Having derived new properties of the copulas (\ref{eq:oldcop})
and (\ref{eq:mixcop}) introduced earlier,
we now seek to generalize them into bivariate models that could be
useful for fitting to data.
The most general form of the bivariate model (\ref{eq:gen}) could be
fitted directly to data for low $n$, and has $(n-1)^2$ parameters.
From the definition of the grade correlation, we have (Nelsen,
2006 \cite{nelsen})
\[\rho_s=12\text{E}(F(x)G(y))-3=12\int\int h(x,y)F(x)G(y)\dee x\dee y-3.\]
Generalizing the proof in Baker \cite{me}, this gives
\begin{equation}\rho_s=\frac{12}{(n+1)^2}\sum_{i=1}^n\sum_{j=1}^n ijr_{ij}-3=\frac{12}{(n+1)^2}\sum_{i=1}^n\sum_{j=1}^n (i-(n+1)/2)(j-(n+1)/2)r_{ij}.\label{eq:rhos}\end{equation}

In the most general case, this model can reproduce any copula
with increasing accuracy as $n \rightarrow \infty$, and all
models of lower order $m < n$ can be written as (\ref{eq:gen})
for some choice of ${\mathbf r}$. However, this does not help in
the construction of simple models with few parameters, which is
our aim. We first discuss several possible approaches, before
introducing what seems the most useful new copula, the `Bessel
function copula'.
\subsection{Generalized bivariate models}
One possibility for reducing the number of model parameters from
$(n-1)^2$ would be to modify the scheme for generating correlated random
numbers given in the introduction, by first generating a
uniformly-distributed random number $U$. Then  a random variable is chosen as
the order statistic number $\lfloor mU \rfloor +1$ of $m$ from $F(x)$, and
number $\lfloor nU \rfloor +1$ of $n$ from $G(y)$, where $\lfloor
\rfloor$ is the `floor' function. This allows the two random variables to be
chosen from different orders of order statistic. The resulting
model is a special case of (\ref{eq:gen}) where $n \ge
m$. Unfortunately, the resulting distributions, characterised by
two discrete parameters,  are not mathematically tractable.

We therefore seek instead to obtain models with only a few
parameters by generalizing (\ref{eq:hn}).  A more general
model arises from pairing order statistics only within some
range or ranges; for example, suppose only the 1st to $m_1$th
and $m_2$th to $n$th order statistics pair, and the remainder
associate randomly. Then
\[H(x,y)=n^{-1}\sum_{k=1}^{m_1}
F_{k,n}(x)G_{k,n}(y)+n^{-1}\sum_{k=n-m_2}^n F_{k,n}(x)G_{k,n}(y)+\frac{\sum_{i=m_1+1}^{n-m_2-1}\sum_{j=m_1+1}^{n-m_2-1}F_{i,n}(x)G_{j,n}(y)}{n(n-m_1-m_2)}.\]
From (\ref{eq:rhos}) it follows that
\begin{multline*} \rho_s=\frac{n-1}{n+1}\\ 
+\frac{2m_1(m_1+1)(2m_1+1)-2(n-m_2-1)(n-m_2)(2n-2m_2-1)}{n(n+1)^2}\\
+\frac{3(n+m_1-m_2-1)^2(n-m_1-m_2)}{n(n+1)^2 }.\end{multline*}

This allows a distribution with three discrete parameters where
the random variables correlate strongly only in one or both tails.

Another way to generate models that are more general than
(\ref{eq:hn}) is to form a finite mixture distribution
\[H(x,y)=\sum_{i=1}^n w_i H^{(i)}(x,y),\]
where $\sum_{i=1}^n w_i=1$. This of course can be expressed
as a special case of (\ref{eq:gen}). The Spearman correlation is
simply
\begin{equation}\rho_s=\sum_{i=1}^n w_i (i-1)/(i+1).\label{eq:spear}\end{equation}

\subsection{The Bessel Function Copula}
A new distribution can be derived by taking an infinite mixture
of models. This gives copulas indexed by one parameter, if the
mixing distribution is a 1-parameter distribution. In
general the pdf is
\begin{equation}h(x,y)=\sum_{n=1}^\infty nw_n\sum_{k=1}^n {n-1 \choose k-1}^2 F(x)^{k-1}(1-F(x))^{n-k}G(y)^{k-1}(1-G(y))^{n-k}f(x)g(y).\label{eq:forrho}\end{equation}
Rearranging,
\[\frac{h(x,y)}{f(x)g(y)}=\sum_{k=1}^\infty
\frac{(F(x)G(y))^{k-1}}{(k-1)!^2}\sum_{n=k}^\infty \frac{n w_n (n-1)!^2\{(1-F(x))(1-G(y))\}^{n-k}}{(n-k)!^2}.\]

An interesting distribution arises on taking
\begin{equation}w_n=\frac{\theta^{n-1/2}}{(n-1)!n!I_1(2\theta^{1/2})},\label{eq:wt}\end{equation}
where $ n > 0$, $\theta > 0$,
and $I$ denotes the Bessel function of imaginary argument. This
is a special case of 2-parameter discrete Bessel function distribution first described by
Pitman and Yor \cite{yor} and later by Yuan and
Kalbfleisch \cite{yuan, kemp}. Then from the series expansion
of the Bessel function, we have that
\begin{equation}\frac{h(x,y)}{f(x)g(y)}=\frac{\theta^{1/2}}{I_1(2\theta^{1/2})}I_0(2(F(x)G(y)\theta)^{1/2})I_0(2\{(1-F(x))(1-G(y))\theta\}^{1/2}).\label{eq:besspdf}\end{equation}

The copula is
\begin{equation}C(u,v)=\frac{\theta^{1/2}}{I_1(2\theta^{1/2})}\int_0^u\int_0^v
I_0\{2(\theta wz)^{1/2}\}I_0\{2(\theta (1-w)(1-z))^{1/2}\}\dee w
\dee z.\label{eq:besscop}\end{equation}
Figures \ref{scatfig1} and \ref{scatfig2} illustrate the copula
as scatterplots, for $\theta=250$ and $\theta=5000$ respectively.
Here a randomly generated sample of size 1000 was generated from
the joint distribution with copula $C$ and uniform marginals.
This copula is the first one known to the author that requires
special functions; all others require only exponentials,
logarithms, and powers.

The Spearman correlation is calculated from (\ref{eq:spear})
and (\ref{eq:wt}) as
\begin{equation}\rho_s=\sum_{n=1}^\infty
\frac{\theta^{n-1/2}}{n!(n-1)!I_1(2\theta^{1/2})}\frac{n-1}{n+1}=\sum_{n=0}^\infty
\frac{\theta^{n+1/2}}{n!(n+3)!I_1(2\theta^{1/2})}=I_3(2\theta^{1/2})/I_1(2\theta^{1/2}).\label{eq:rhos1}\end{equation}
To obtain negative correlations, one sets $G(y)
\rightarrow 1-G(y)$.

As $\theta \rightarrow 0$, (\ref{eq:besspdf}) gives $h(x,y)
\rightarrow f(x)g(y)$. As $z \rightarrow \infty$, since
$I_\nu(z) \rightarrow \exp(z)/\sqrt{2\pi z}$, we have that as
$\theta \rightarrow \infty$
\[\frac{h(x,y)}{f(x)g(y)} \sim
\frac{\exp(-2\theta^{1/2}T(x,y))}{2(\pi)^{1/2} \theta^{1/4}\{F(x)(1-F(x))\}^{1/4}\{G(y)(1-G(y))\}^{1/4}},\]
where
$T(x,y)=(F^{1/2}(x)-G^{1/2}(y))^2+((1-F(x))^{1/2}-(1-G(y))^{1/2})^2$.
This shows that $h(x,y) \rightarrow 0$ if $F(x) \ne G(y)$.
Hence as $\theta \rightarrow \infty$ the distribution attains
the Fr\'{e}chet bound. From (\ref{eq:rhos1}) as $\theta
\rightarrow \infty$, we have that $\rho_s \rightarrow 1$, so the
grade correlation approaches unity, as it must.

As this copula is not a finite mixture of the copula
(\ref{eq:oldcop}), the coefficient of tail dependence could be
nonzero. However, using the reflection symmetry of the copula,
we have that the coefficient of right (and left) tail dependence is
$\lim_{p\rightarrow 0}C(p,p)/p$, where the double integral in
(\ref{eq:besscop}) is $O(p^2)$. The coefficient of tail
dependence is thus still zero, except of course in the limit as the
Fr\'{e}chet-Hoeffding bound is approached as $\theta\rightarrow\infty$.

Random variables from this copula can be derived by generating
$N$ from the discrete Bessel distribution, as described by
Devroye \cite{devroye}, and then randomly selecting one of the $N$
order-statistic pairs. This is how figures \ref{scatfig1} and
\ref{scatfig2} were generated. Here, $N$ was generated using the
inverse probability method. This general strategy would be efficient if many
random numbers were required, when unused order statistic pairs
could be stored and used in preference to generating fresh ones.
It also requires only generation of random numbers from the
marginal distributions, and does not require the use of the
inverse probability transformation on these distributions. The
alternative method, of generating $U$ and then generating $V$
from the conditional distribution $\partial C(u,v)/\partial u$
is not recommended as it is computationally more time consuming.

One can also take the weight 
\begin{equation}w_n=\frac{\theta^{n-1}}{(n-1)!^2I_0(2\theta^{1/2})},\label{eq:wt0}\end{equation}
another special case of the Bessel function distribution.
After summing the series, this yields the more complex form
\[\frac{h(x,y)}{f(x)g(y)}=\frac{AI_1(2A)I_0(2B)+BI_0(2A)I_1(2B)+I_0(2A)I_0(2B)}{I_0(2\theta^{1/2})},\]
where $A=\{F(x)G(y)\theta\}^{1/2}$, $B=\{(1-F(x))(1-G(y))\theta\}^{1/2}$.
The Spearman correlation from (\ref{eq:spear}) is
\[\rho_s=\{2\theta^{-1/2}I_3(2\theta^{1/2})+I_4(2\theta^{1/2})\}/I_0(2\theta^{1/2}).\]
The copula is in general similar to (\ref{eq:besscop}) but
slightly more complex.

Still other choices can be made for $w_n$, but these lead to pdfs that
are much less tractable, being infinite sums of hypergeometric
functions. The Spearman correlations however are more tractable; for
example taking
a displaced Poisson distribution for the weights
$w_n=\theta^{n-1}\exp(-\theta)/(n-1)!$, the Spearman correlation
may be shown to be
\[\rho_s=1-2\theta^{-1}+2\theta^{-2}(1-\exp(-\theta)).\]

\section{Bivariate Data Fitting Example}
\subsection{Australian Institute of Sports data}
A dataset from the Australian Institute of Sport is used as an
example.  This is given in Cook and Weisberg \cite{cook} and has
been used as a testbed for new distributions by Azzalini and
others \cite{azz1999,azz2003}. Here, percentage body fat and weight of
102 male athletes were used. Figure \ref{fig2} shows the skew
distribution of percentage body fat. The distribution of weight
(not shown) was also slightly skew.

A suitable univariate model for the marginal distributions was
chosen as the lagged normal distribution \cite{davis}, where the
random variable $X=Z+Y$, where $Z$ is Gaussian, and $Y$ is
exponential. In fact, taking $X=Z+Y_1-Y_2$ gives a distribution
that can be skew in either direction and long-tailed to either
or both left and right. Taking the
normal mean as $\xi$ and standard deviation $\beta$, and the
exponential means as $\alpha_1$ and $\alpha_2$, the pdf is
\begin{multline}f(x)=\frac{1}{\alpha_1+\alpha_2}[\exp\{\frac{1}{2}(\beta/\alpha_1)^2-(x-\xi)/\alpha_1\}\Phi(\frac{x-\xi}{\beta}-\frac{\beta}{\alpha_1})\\
+\exp\{\frac{1}{2}(\beta/\alpha_2)^2+(x-\xi)/\alpha_2\}\Phi(-\frac{x-\xi}{\beta}-\frac{\beta}{\alpha_2})],\label{eq:pdflag}\end{multline}
where $\Phi$ is the normal distribution function. The
distribution function $F(x)$ is
\begin{multline}F(x)=\Phi(\frac{x-\xi}{\beta})+\frac{1}{\alpha_1+\alpha_2}[-\alpha_1\exp\{\frac{1}{2}(\beta/\alpha_1)^2-(x-\xi)/\alpha_1\}\Phi(\frac{x-\xi}{\beta}-\frac{\beta}{\alpha_1})\\
+\alpha_2\exp\{\frac{1}{2}(\beta/\alpha_2)^2+(x-\xi)/\alpha_2\}\Phi(-\frac{x-\xi}{\beta}-\frac{\beta}{\alpha_2})].\label{eq:cdflag}\end{multline}
The mean $\mu=\xi+\alpha_1-\alpha_2$, variance
$\sigma^2=\beta^2+\alpha_1^2+\alpha_2^2$, skewness
$\gamma=2(\alpha_1^3-\alpha_2^3)/\sigma^3$, and kurtosis $\kappa=6(\alpha_1^4+\alpha_2^4)/\sigma^4$.
Since $\Phi$ is a well-known special function, and the
distribution function can be written as a function of $\Phi$,
and also the moments can be written down, this distribution is
quite an attractive choice for fitting data that depart from
normality, and are not heavy tailed. The easy computation of the
distribution function makes it particularly attractive for use
in fitting multivariate distributions via copulas. Care is
needed in computing the pdf and distribution function when
$\alpha_1/\beta$ or $\alpha_2/\beta$ are small. One can then use the
asymptotic expansion for $\Phi(z)$, 
\[\Phi(z)=\frac{\exp(-z^2/2)}{(2\pi)^{1/2}}\{-1/z+1/z^3-3/z^5+3
\times 5/z^7-\cdots\},\]
which avoids the rounding errors implicit in taking the product
of  very large and  very small quantities.
The symmetric form of this distribution, with
$\alpha_1=\alpha_2$, is described in Johnson, Kotz and
Balakrishnan \cite{univ}, vol. 2, chap. 24.

The percentage body fat could be fitted by maximum likelihood to
a lagged normal distribution, where only the right tail was
needed, so that $\alpha_2=0$.  Figure \ref{fig2} shows the
fitted curve, with Azzalini's skew normal distribution
\cite{azz2003} also
fitted. Both distributions fitted satisfactorily, according to
the Kolmogorov test, although better fits can be achieved at the
expense of using more parameters; there is even a suggestion of
bimodality in the data. This is possible, as the sample
comprises athletes from a variety of different sports.
Weight and height look normal, but weight has a lower AIC
(Akaike Information Criterion) if fitted to
a lagged normal, and this was done.

The bivariate pdf (\ref{eq:pdfp}) with $n=10$ fitted the
data with a log-likelihood of $\ell=-607.54$ and a weight
$q=0.78$. The observed Spearman and Pearson correlations were
0.613 and 0.581, and the predictions from the model were  0.640
and 0.576. The Bessel function pdf (\ref{eq:besspdf}) also
fitted satisfactorily, with $-\ell=606.47$, and predicted
Spearman correlation of 0.65, Pearson correlation 0.565. The
fitted value of $\theta$ was $\hat{\theta}=23.7$. For
comparison, the Azzalini bivariate distribution \cite{azz2003}
fitted with $-\ell=612.22$, with the same number (7) of
parameters. The point here is that the Bessel function copula (\ref{eq:besscop})
performs satisfactorily, as does the whole copula-based
methodology of modeling the marginal distributions
individually, and gluing them together with a copula.  One can
obtain good fits to the data, without forcing both the marginal
distributions to be of the same form. There is then the freedom
to vary the marginal modeling, for example by fitting a bimodal
distribution in figure \ref{fig2}, which option is not available on
fitting a standard multivariate distribution.

\section{Multivariate models}
Consider the multivariate generalization of the models presented
so far. This topic was only briefly touched on in \cite{me},
and the results here are new.

Denote the $i$th of $p$ random variables by $X_i$, and denote
the corresponding distribution functions, pdfs, and distribution
functions of the $k$th of $n$ order statistics by
$F^{(i)}(x_i), f^{(i)}(x_i)$ and $F^{(i)}_{k,n}(x_i)$
respectively.  The most general multivariate model of order $n$
would be
\[H({\mathbf x})=\sum_{k_1\cdots k_p}\prod_{i=1}^p
F_{k_i,n}^{(i)}(x_i)r_{k_1\cdots k_p} ,\]
where $r_{k_1\cdots k_p} \ge 0$ and $\sum_{k_j, j \ne
m}r_{k_1\cdots k_p} =1/n$ for all $m$.

This would have $(n-1)^p$ parameters. To reduce this number, one
could consider only models in which the random variables are in phase (or in
antiphase, for negative correlations) in cycles of length $n$.
Variables could all be in the same cycle, or some could be in
independent cycles. For example, with 5 variables, two could be
paired in one cycle, two in another independent cycle, and the
fifth variable be in a cycle of its own. To generate random numbers
from such a distribution, one could compute the $n$ order
statistics for the 5 variables, and then one random number would
decide which order statistic was to be taken for variables 1 and
2, another independent random number would select the third and
fourth random variable pair, and a third independent random choice would
select the fifth random variable from among its $n$ order statistics.
Clearly, random numbers for variables in such single cycles
could be more efficiently generated by simply choosing a random variable from
the appropriate marginal distribution.

The number of models $a_p$ is the number of ways $p$
distinguishable objects (random variables) fit into $p$ or fewer identical
boxes (cycles). This is given by the recursion relation
\[a_p=\sum_{j=0}^{p-1}{p-1 \choose j}a_{p-1-j},\]
with $a_0=1$ (Tucker, \cite{tucker}). This may be derived by considering
the addition of the $p$th object. It must occur in a box
containing $0 \le j \le p-1$ other objects, where the $j$ other objects
can be chosen in ${p-1 \choose j}$ ways, and the remaining
$p-1-j$ other objects in the other boxes can be arranged in
$a_{p-1-j}$ ways. The recursion relation follows, and the number
of mixing parameters for a mixture model is $a_p-1$. Table 1
shows the number of models resulting; the number grows faster
than exponentially with $p$. The table also shows the number of
parameters $2^p-p-1$ for the subset of models  obtained by simply
including or excluding random variables from one common cycle. This simple
scheme gives distributions whose marginals allow differing
Spearman correlations, and is feasible up to dimensions of 5 or
6, beyond which the number of model parameters becomes
excessive. 
The multivariate distribution function $H({\mathbf x})$ can be written
\begin{equation}H({\mathbf x})=\sum_{S=1}^{2^p}w_S\prod_{i \notin
S}F^{(i)}(x_i)\{n^{-1}\sum_{k=1}^n\prod_{j \in S}F_{k,n}^{(j)}(x_j)\},\label{eq:bigh}\end{equation}
where the sets $S$ run through all possible subsets of the $p$
random variables, and where
$\sum_{S=1}^{2^p}w_S=1$.
The model parameters can be estimated in the same way
as described earlier for bivariate mixture models.

\section{Multivariate Data Fitting Examples}
With the notation
\[T_{ij}(x_i,x_j)=\frac{1}{n}\sum_{k=1}^n F_{k,n}^{(i)}(x_i)F_{k,n}^{(j)}(x_j)\]
etc, the trivariate case of (\ref{eq:bigh}) can be written
\begin{multline}H({\mathbf
x})=w_0F^{(1)}(x_1)F^{(2)}(x_2)F^{(3)}(x_3)+w_1F^{(1)}(x_1)T_{23}(x_2,x_3)+w_2F^{(2)}(x_2)T_{13}(x_1,x_3)\\
+w_3F^{(3)}(x_3)T_{12}(x_1,x_2)+w_4T_{123}(x_1,x_2,x_3),\label{eq:3}\end{multline}
where $w_0+w_1+w_2+w_3+w_4=1$.

This model was fitted to the percentage of body fat, weight, and
height for the Australian Institute of Sport data from Cook and
Weisberg \cite{cook}. 

After fitting the three marginal distributions by maximum
likelihood, using the lagged normal distribution, the
trivariate distribution was fitted by maximum likelihood to find
the four parameters $w_1 \ldots w_4$, keeping the marginal
distributions fixed. Subsequently, allowing the parameters of
the marginal distributions to float increased the log-likelihood
$\ell$ by only a very small amount.  Fitting weights that sum to
unity poses a computational problem.  The simple solution
adopted was to use parameters $v_0$ to $v_4$, fixing one
parameter $v_4=0$, and then the $w_i=\exp(v_i)/\sum_{j=0}^4
\exp(v_j)$ sum to unity, while the 4 free parameters can take
any value on the real line. Any of the $v_j$ can be set to 0,
but the choice is best altered if the term chosen fits to very
small weight $w_j$, as then all the other $v_j$ become huge.

The observed and predicted Pearson and Spearman correlations are
given in table \ref{t:two}, showing fairly good agreement.
A value of $n=12$ was used, but the results are not very
sensitive to this, as long as $n$ is large enough to allow the
highest correlation.
The five fitted weights in (\ref{eq:3}) were respectively $ 0.0003,  0.435,
0.0112, 0.284,  0.270$. Interestingly, the independence (first) term is
not needed. A common Spearman correlation of $(11/13) \times
0.27$ derives from the last term in (\ref{eq:3}), and the
correlation between  weight and height is then boosted by the
second term, while that between body fat and weight is boosted
by the fourth term.

The trivariate fit by this copula fits just slightly worse than
the 3-dimensional Azzalini model, in terms of log-likelihood,
even although the marginal fits are slightly better.
The Azzalini model gave $\ell=-931.3$, while this model gave $\ell=-932.6$.
Unfortunately, this is the `little rift within
the lute' that limits the usefulness of these multivariate
models. Although they can accommodate large and variable
correlations, they cannot fit an arbitrary correlation matrix.
This was seen much more clearly on moving to a quadrivariate
example, taken from a study by Penrose \cite{penrose} and
available online via Statlib, in which
percentage body fat, weight, height and abdominal circumference
were fitted, for a sample of 252 men. 

The quadrivariate model in a terse notation, writing e.g. $F^{(1)}(x_1)=T_1$, was 
\begin{multline*}H({\mathbf x})=w_0T_1T_2T_3T_4\\
+w_1T_1T_2T_{34}+w_2T_1T_3T_{24}+w_3T_1T_4T_{23}
+w_4T_2T_3T_{14}+w_5T_2T_4T_{13}+w_6T_3T_4T_{12}\\
+w_7T_1T_{234}+w_8T_2T_{134}+w_9T_3T_{124}+w_{10}T_4T_{123}\\
+w_{11}T_{1234}+w_{12}T_{12}T_{34}+w_{13}T_{13}T_{24}+w_{14}T_{14}T_{23}.
\end{multline*}

Although the lagged normal
distribution gave satisfactory marginal fits, the quadrivariate
model fitted with 14 parameters gave $-\ell=3352$, compared with
the quadrivariate Azzalini distribution fit of $-\ell=3184$, and
most of the weights fitted as zero.
It was clear that the fitted correlations were in general too
small.  Hence the usefulness of these multivariate distributions
seems limited.
\section{Conclusions}
Following the introduction of a new copula in Baker \cite{me}, it
became clear to the author that its properties had not been fully enumerated,
and also that it was possible to derive further copulas by
generalizing it. Further, only a few bivariate distributions had been
fitted to data, and there was no practical experience at all with fitting
multivariate distributions.

In this paper, several ways of extending the class of copulas
have been given. Perhaps the most promising one is to make the
order $n$ a random variable from the discrete Bessel distribution. This leads to the `Bessel function'
copula (\ref{eq:besscop}), the only copula in the author's
experience that requires a special function for its
expression. This copula is indexed by one parameter $\theta$.
Like the Frank, Clayton and Plackett copulas, it contains the
independence case $C(u,v)=uv$, and can attain the Fr\'{e}chet
bound as $\theta \rightarrow\infty$. Negative correlations are
dealt with by e.g.  setting $G(y) \rightarrow 1-G(y)$. The use
of this copula has been illustrated by fitting it to the
Australian Institute of Sport dataset \cite{cook}. The fact that
the copula must be written either as a double integral, or as a
series expansion is a drawback, but in fitting to data by
likelihood-based methods, the crucial requirement is that the
pdf must be easily computable. This pdf is easy to compute,
given the widespread existence of routines to compute the
special functions $I_0$ and $I_1$. It is also not difficult to
generate random variables. This copula is by the way not
Archimedean; Archimedean copulas must be associative, but
computations showed a difference between $C(u,C(v,w))$ and
$C(C(u,v),w)$ (lack of associativity) of up to about 2\%.

The properties of the bivariate copulas have been further
explored. The most significant is probably that the original
copula in (\ref{eq:plus}) and its mixtures possess the LRD
(likelihood ratio dominance) ordering property. This property is
therefore also possessed by the Bessel function copula. 

The properties of the analogous multivariate copulas have also
been studied, but here results are less positive.  They do have
some flexibility; marginal distributions need not have identical
parameters, and high correlations can be accommodated. Although
the hitherto untried process of fitting trivariate and
quadrivariate models to data by maximum-likelihood estimation
proved entirely feasible, it seems that despite their many
parameters these distributions can not reproduce an arbitrary
correlation matrix. The use of these distributions for $p > 2$
is therefore problematical. They may however prove to be a
starting point for the development of more useful distributions.
\clearpage
\begin{figure}[ht]
\centering
\makebox{\includegraphics{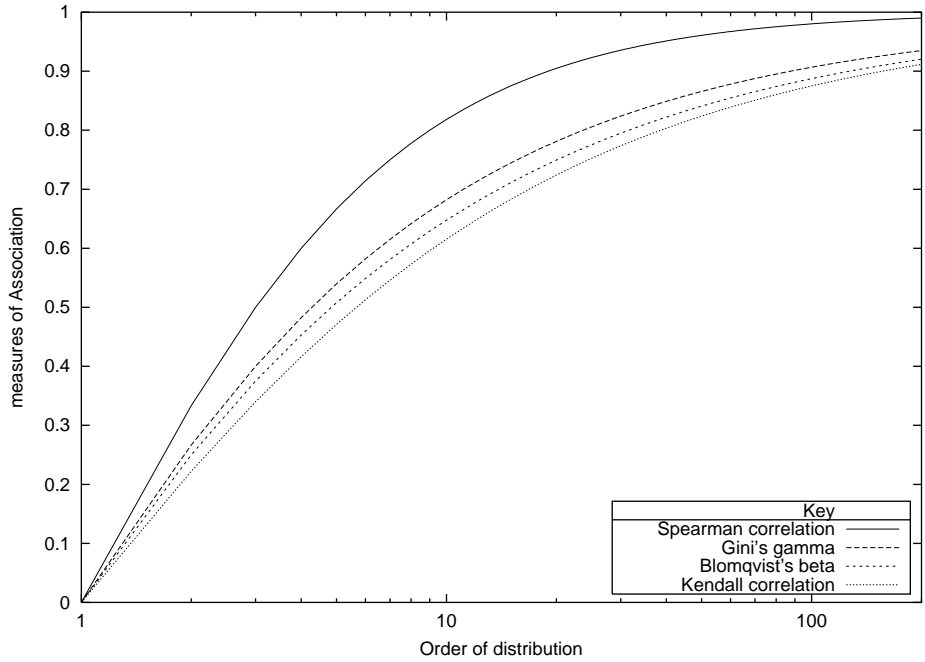}}
\caption{\label{fig1}Measures of association as a
function of the order $n$ of the distribution for the
distribution of equation (\ref{eq:hn}). The key gives the curves
from top to bottom.}
\end{figure}

\begin{figure}[ht]
\centering
\makebox{\includegraphics{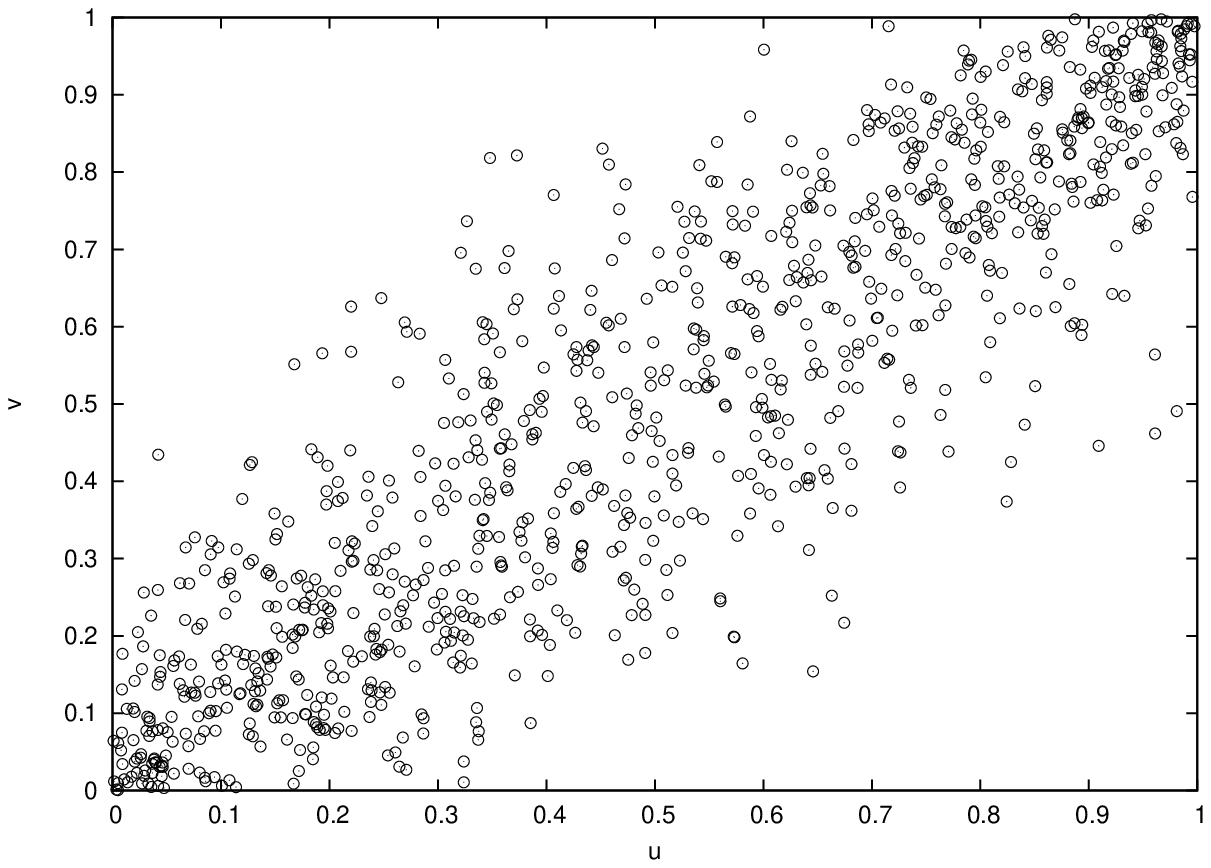}}
\caption{\label{scatfig1}The Bessel function copula with $\theta=250$.}
\end{figure}

\begin{figure}[ht]
\centering
\makebox{\includegraphics{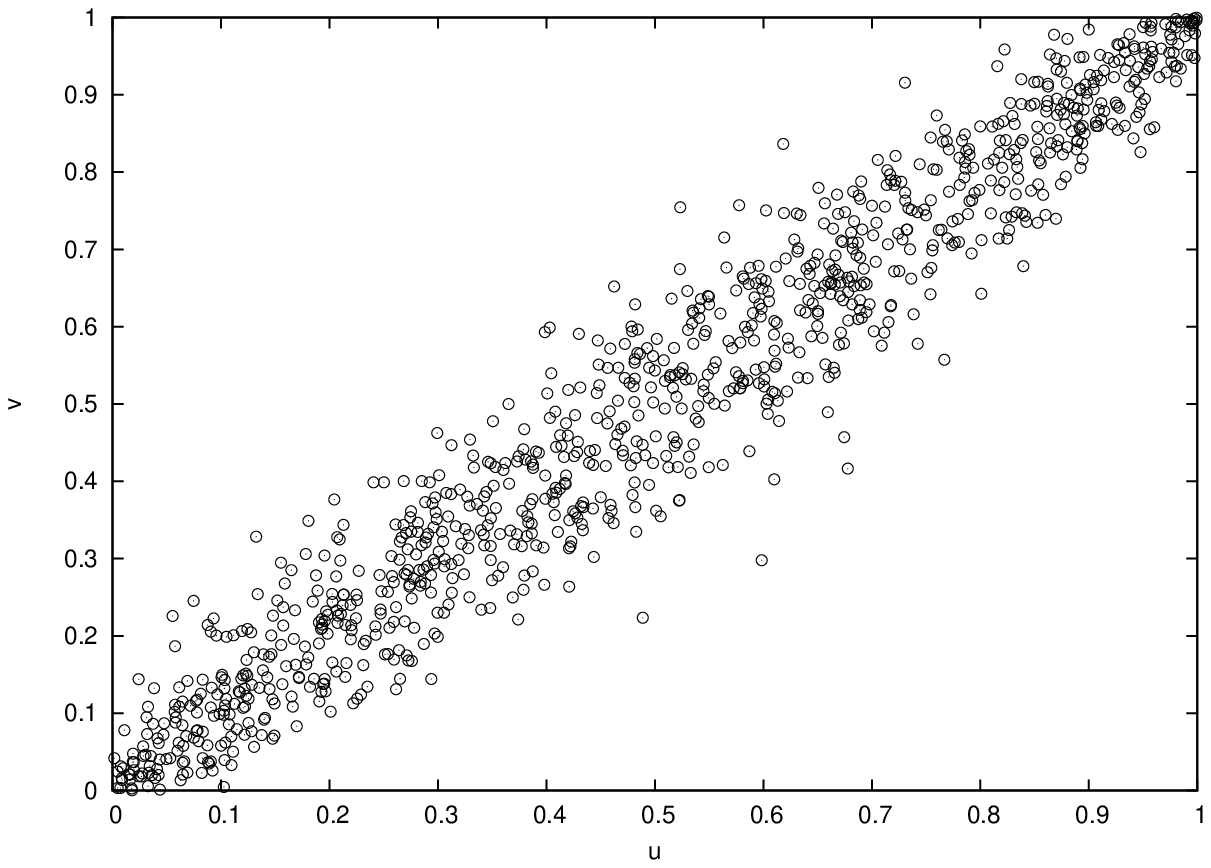}}
\caption{\label{scatfig2}The Bessel function copula with $\theta=5000$.}
\end{figure}

\begin{figure}[ht]
\centering
\makebox{\includegraphics{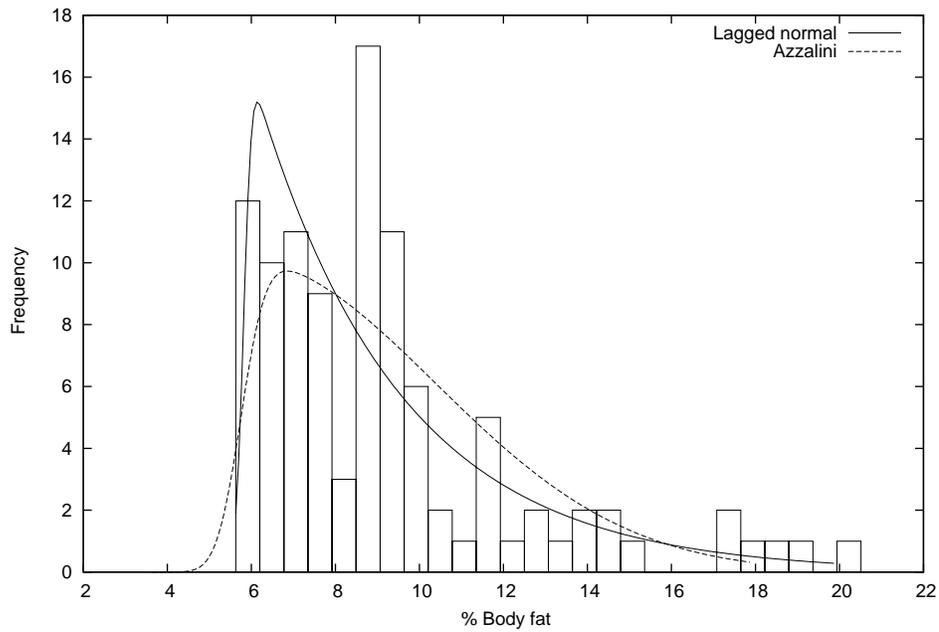}}
\caption{\label{fig2}Fits of the lagged normal and Azzalini
distributions to the percentage of body fat for 102 male athletes
(Australian Institute of Sport data).}
\end{figure}
\begin{table}
\caption{Numbers of parameters to be estimated for two
classes of multivariate model, and the number of correlations.
The models, from left to right, are the 
single cycle model, and the multicycle model.}
\label{t:one}
\begin{center}
\begin{tabular}{llll} \hline \hline
Dimension $p$ & Params, model 1 & Params, model 2 & $p(p-1)/2$\\ \hline
2 & 1 & 1 & 1 \\ \hline
3 & 4 & 4 & 3 \\ \hline
4 & 11 & 14 & 6 \\ \hline
5 & 26 & 51 & 10 \\ \hline
\end{tabular}
\end{center}
\end{table}

\begin{table}
\caption{Observed and predicted Pearson correlations ($\rho$)
and Spearman correlations ($\rho_s$) on fitting the trivariate
model in equation \ref{eq:3} with lagged normal marginals.}
\label{t:two}
\begin{center}
\begin{tabular}{lllll} \hline \hline
Variables  & Obs. $\rho$ & Pred. $\rho$ & Obs. $\rho_s$ & Pred. $\rho_s$\\ \hline
\% Body fat \& Weight &  0.581 & 0.412 & 0.613 & 0.468\\ \hline
\% Body fat \& Height &  0.192 & 0.199 & 0.237 & 0.237\\ \hline
Height \& Weight &  0.666 & 0.596 & 0.677 & 0.596 \\ \hline
\end{tabular}
\end{center}
\end{table}
\clearpage

\end{document}